\documentclass[10pt,a4paper]{article}
\usepackage[margin=0.75in]{geometry}
\usepackage{amsbsy,amsfonts,amsmath,amssymb,color,epsfig,graphicx,natbib,rotating,subeqn,subfigure}

\usepackage{epsfig,epic,eepic,color,rotate,graphics,fancybox}
\usepackage{graphicx}
\usepackage{epstopdf}
\usepackage{color}
\usepackage{bm}
\begin{document}
\title{Taking one charge off a two-dimensional Wigner crystal}
\author{Moritz Antlanger$^{a,b}$$^{\ast}$\thanks{$^\ast$Corresponding author. Email: moritz.antlanger@tuwien.ac.at}, Martial Mazars$^{b}$, Ladislav \v{S}amaj$^{c,d}$, Gerhard Kahl$^{a}$, and Emmanuel Trizac$^{c}$\\
$^{a}${\em \footnotesize Institute for Theoretical Physics and Center for Computational Materials Science (CMS), Vienna University of Technology, Wien, Austria} \\
$^{b}${\em \footnotesize Laboratoire de Physique Th\'eorique (UMR 8627), Universit\'e Paris-Sud and CNRS, Orsay, France} \\
$^{c}${\em \footnotesize Laboratoire de Physique Th\'eorique et Mod\`eles Statistiques (UMR 8626), Universit\'e Paris-Sud and CNRS, Orsay, France}\\ 
$^{d}${\em \footnotesize Institute of Physics, Slovak Academy of Sciences, Bratislava, Slovakia} \\
\vspace{6pt} 
{\bf preprint: LPT-ORSAY-14-10}
\vspace{6pt}
}
\maketitle

\begin{abstract} 
A planar array of identical charges at vanishing temperature forms a
Wigner crystal with hexagonal symmetry.  We take off one (reference)
charge in a perpendicular direction, hold it fixed, and search for the
ground state of the whole system.  The planar projection of the
reference charge should then evolve from a six-fold coordination
(center of a hexagon) for small distances to a three-fold arrangement
(center of a triangle), at large distances $d$ from the plane.  The
aim of this paper is to describe the corresponding non-trivial lattice
transformation. For that purpose, two numerical methods (direct energy
minimization and Monte Carlo simulations), together with an analytical
treatment, are presented. Our results indicate that the $d=0$ and
$d\to\infty$ limiting cases extend for finite values of
  $d$ from the respective starting points into two sequences of stable
  states, with intersecting energies at some value $d_t$; beyond this value the
  branches continue as metastable states.
\end{abstract}

\bigskip


\section{Introduction}
\label{introduction}

In 1934, Wigner \cite{Wigner34} pointed out a possible crystallization
of a three-dimensional (3D) quantum jellium (one-component plasma),
consisting of charged particles immersed in a homogeneous neutralizing
background, at low densities. The possibility of the formation of a
two-dimensional (2D) crystal of electrons on the surface of liquid
helium and in inversion layers of semiconductors at low temperatures
was predicted theoretically in Refs. \cite{Crandall71} and
\cite{Chaplik72}, respectively.  The corresponding experiments of an
electron gas trapped at the surface of liquid helium was realized by
Grimes and Adams \cite{Grimes79}, in the semiconductor structure
GaAs/GaAlAs by Andrei {\it et al.} \cite{Andrei88}, and of
laser-cooled $^9$Be$^+$ ions confined in Penning traps by Mitchell
{\it et al.} \cite{Mitchell:98}. For reviews about classical and
quantum Coulomb crystals, see
e.g. \cite{Bonitz08,Monarkha12,Monarkha:book:03}.

From a theoretical point of view, the ground-state energies of a
classical 2D electron crystal and the phonon spectra were studied for
a variety of Bravais lattices in Refs.  \cite{Platzman74,Bonsall77},
with the conclusion that a simple hexagonal structure (built up by
equilateral triangles) provides the lowest energy.  To understand the
thermodynamics and the dynamical properties of electrons at low
temperatures, deviations from a perfect crystal have been studied in
the seminal work of Fisher {\it et al.}  \cite{Fisher79}.  These
investigations involve (i) localized low-energy defects (such as
vacancies, interstitials, etc.) which are expected to govern dynamical
properties of migrating electrons, and (ii) extended defects with
higher energies (such as dislocations, grain boundaries, etc.) which
are supposed to play an important role in the melting process of the
crystal.  At the present stage of knowledge, grain boundaries are
responsible for melting 3D Wigner crystals, while the
Kosterlitz-Thouless theory of dislocations and disclinations
\cite{Kosterlitz73,Nelson79} describes the melting of 2D electron
crystals.
Related models involve curved geometries \cite{Messina1,Messina2},
large 2D Coulomb clusters confined by a harmonic potential
\cite{Bedanov94,Kong03}, 2D colloidal crystals with pair interactions
of Yukawa \cite{Libal07,He13} or $1/r^3$ \cite{Lechner09,Teeffelen13}
forms.

\begin{figure} 
\begin{center}
\includegraphics[width=.8\linewidth]{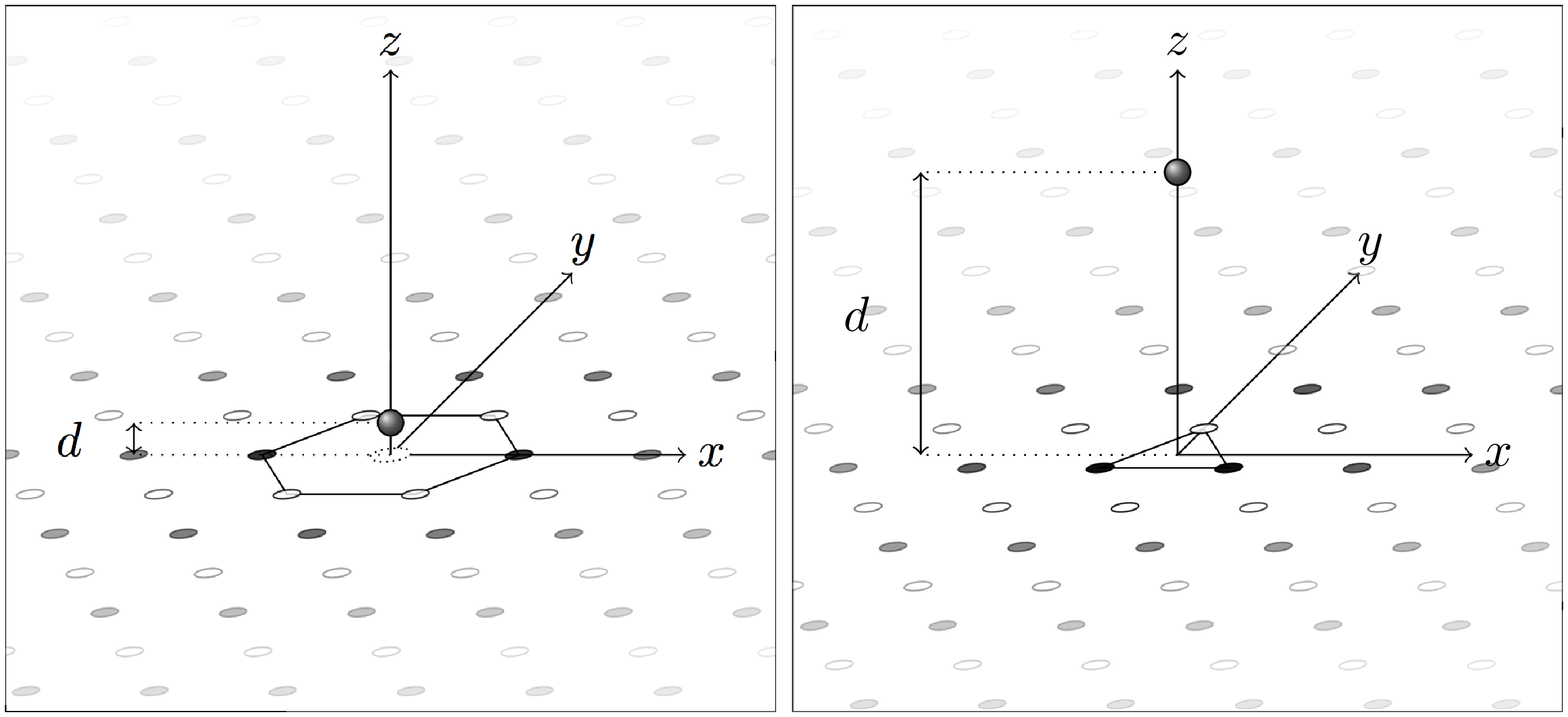} 
\end{center} 
\caption{Schematic representation of the two limiting configurations
  expected.  Left panel: When the reference charge is close to the
  plane, its perpendicular projection is endowed with six-fold
  coordination. Right panel: at large distances $d$, its coordination
  number becomes three.  In the remainder, the dimensionless
  counterpart of $d$ will be denoted $\eta$.}
\label{sketch}
\end{figure}

In the present paper, we study from a classical perspective the
ground-state problem of taking off a charge from a bidimensional
crystal.  Our starting point is a perfect, 2D Wigner crystal which we
assume to be embedded in the $(x, y)$-plane. It is formed by particles
(each with a negative elementary charge of $-e$), localized at the
sites of a hexagonal lattice (with lattice spacing $a$). The charges
of the particles are neutralized by a uniform background of charge
density $\sigma e$. Then, we take one of the charges (carrying the
index $0$, and coined the ``reference'' or ``tagged'' particle) away from the crystal and fix it at
a distance $d$ in the vertical $z$-direction. As we consider
increasingly large values for $d$, the remaining particles in the
Wigner crystal will leave their original, regular lattice positions
and will intuitively approach the vacancy left behind by the tagged
particle.  This spatial deformation is realized in an effort to
minimize the total interaction energy of the setup (i.e., ``Wigner
crystal + tagged charge''). It stands to reason that the removal of
charge 0 has a stronger impact on the particle positions of the
lattice the closer these particles have been to the tagged charge in
the original Wigner crystal (i.e. at $d=0$). However, one should not
forget that due to the long-range nature of the Coulomb potential, the
interactions between all particles are important.  Two limiting cases
can be envisioned.  (i) When $d=0$, and presumably when $d/a \ll 1$,
the reference particle has coordination six, see Figure \ref{sketch}.
(ii) On the other hand, at asymptotically large distances (i.e., for
$d/a \gg 1$), the Wigner crystal is interacting only weakly with the
tagged particle and therefore the perfect hexagonal structure of the
lattice should be maintained. Under these conditions, the total
optimal configuration is realized when the projection of the reference
particle coincides with the center of any triangle formed by three
neighbouring particles of the Wigner crystal (see right panel of
Figure \ref{sketch}).  The transformation of the Wigner crystal from a
six-fold coordination (valid at least for $d=0$) to a three-fold
coordination (valid at least for $d\to \infty$), induced by a change
in the distance $d$ of the tagged particle, represents the central
topic of this contribution.

To the best of our knowledge, the proposed problem has not been
addressed so far. It naturally appears when studying the
strong-coupling regime of counter-ions close to a uniformly charged
wall \cite{Netz01,Samaj11}.  It also is of relevance within the
context of phenomena such as evaporation of particles from a surface
at low temperatures or the creation of lattice defects by manipulating
individual particles \cite{Pertsinidis01a,Pertsinidis01b}. Several
fundamental questions arise:

\begin{itemize} 
\item[(i)] Does the six-fold coordination of the tagged charge change
  to a three-fold coordination (or even to some other value) at a {\it
    finite} distance $d_t$ or already at an infinitesimally small
  value?
\item[(ii)] What is the nature of the transition as the six-fold
  coordination is lost? In particular, is it {\it continuous}, i.e. do
  the six nearest neighbours of the tagged particle rearrange in a
  continuous fashion into some non-equivalent subsets of particles,
  each specified by a different shift away from their original crystal
  positions? Or is the transition {\it discontinuous}, accompanied by
  a change in the {slope of the energy} at the transition distance,
  $d_t$?
\item[(iii)] If the three- and six-fold coordinated limiting states
  lead to metastable configurations at finite $d$, what is the
  corresponding energy barrier? Does it take a finite value, or does
  it scale with the number of particles, $N$?
\end{itemize}

The last question is relevant in view of practical realization of the
``experiment'', and also pertinent for computational purposes. Our
analysis will show that metastable states can coexist for all
distances, separated by an energy barrier that seems high
enough so that the system will stay in a local energy minimum also
after crossing the transition distance $d_t$.  In that case, when
increasing $d$, one observes a hysteresis similar to that of
ferromagnetic systems.  In the ferromagnetism of two macroscopic $+$
and $-$ magnetized states, one needs a relatively large opposite
magnetic field to reverse the magnetization of a macroscopic domain.
In our problem, the role of the magnetic field is taken over by the
distance $d$: if $d \gg d_t$, the local minimum (a reminiscence of the
six-fold coordinated state at $d=0$) might become unstable, or might start to transform into a precursor of the state
with three-fold coordination.

The metastability feature will lead us to define two branches (see
Section \ref{model}): the ``out''-branch (extrusion) where $d$
increases, starting from 0 where the coordination is six-fold, and the
``in''-branch (intrusion), starting from large $d$ where the
coordination number is three, and decreasing $d$ down to 0. It should
be kept in mind that we shall consider a sequence of equilibrium
situations only, at vanishing temperature (i.e. we let the system find the lowest total energy
configuration, for every $d$).  In doing so, we will answer a few of
the questions addressed above. We have used three theoretical tools:
energy minimization, Monte Carlo simulations (both methods being
purely numerical), and an analytic approach. They all
bear their own limitations, since simplifying assumptions were made to
allow for solutions: in both numerical approaches we have considered a
unit cell (containing a sufficient number of particles) that contains
a finite section of the Wigner crystal as well as the tagged charge;
for a fixed position of charge 0, all particles of the remaining
lattice are allowed to freely relax their position [remaining in the
  same $(x,y)$ plane]. The entire system is then a periodic
replication of this unit cell. In the analytic approach, the system is
assumed to be of infinite extent in the $(x, y)$-direction; however,
spatial relaxations were allowed only for the nearest neighbours of
the tagged charge, for simplicity.

The manuscript is organized as follows. The model is specified in
detail in Section \ref{model}, where the two branches are introduced.
We then present in Section \ref{numerical-approach} the basic features
of our two numerical approaches, energy minimization and Monte Carlo
simulations. Both methods rely on Ewald summation techniques, to take
due account of the long range nature of the interaction potential.
The analytic approach is presented in Section \ref{analytic-approach},
and the results are subsequently discussed in Section \ref{results}.
The paper closes with our conclusions and outlook on future work in
Section \ref{conclusion}.  An Appendix collects cumbersome expressions
required for the analytic treatment.

\section{The system and the two branches protocol}
\label{model}

\subsection{Definition of the model}

\begin{figure}
\begin{center}
\includegraphics[width=0.4\textwidth,clip]{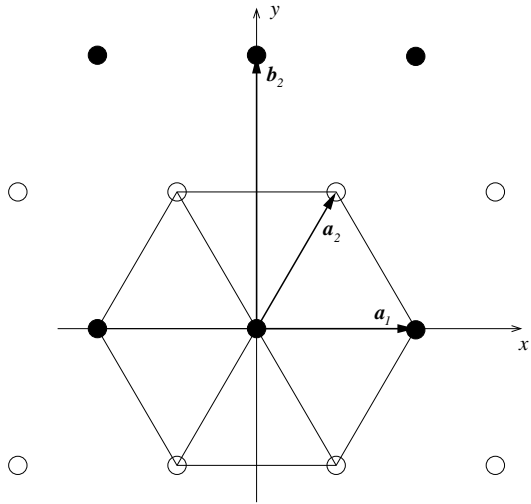}
\caption{Hexagonal structure of the undistorted 2D Wigner crystal, as
  obtained for $d=0$ or $d\to\infty$.  The arrows show
  $\bm{a}_1=\bm{b}_1$ and $\bm{a}_2$, the primitive lattice vectors
  defined in Equation (\ref{lattice_hexagonal}), while $\bm{b}_2$ is
  defined in Equation (\ref{eq:defb}). Black and white colors are for
  alternating rows.}
\label{hexagonal} \end{center}
\end{figure}

We start from the hexagonal structure of the 2D Wigner crystal: its
unit cell is a rhombus defined via the primitive lattice vectors

\begin{equation}
{\boldsymbol a}_1 = a (1,0) , \qquad 
{\boldsymbol a}_2 = \frac{a}{2} (1,\sqrt{3}) ,
\end{equation}
where $a$ is the lattice spacing (see Figure \ref{hexagonal}). The
positions in the 2D lattice,
\begin{equation}
\label{lattice_hexagonal}
{\bf R}_j = (R_j^x,R_j^y) = j_1 \, \bm{a}_1 + j_2 \, \bm{a}_2 
\end{equation} 
are indexed by $j=(j_1,j_2)$, where $j_1$ and $j_2$ are {arbitrary
  integers. Due to the single-occupancy of our crystal, we can use $j$
  as the particle index.}  There exists another, {equivalent}
representation of the hexagonal structure.  Let us {label the
  particles along rows} alternately
{with} white and black colour.  {In doing so, we obtain} two
{identical, rectangular} sub-lattices ('white' and 'black'), {each of
  them defined via} orthogonal translational vectors
\begin{equation}
\bm{b}_1 \equiv \bm{a}_1 = a (1,0) , \qquad 
\bm{b}_2 = a (0,\sqrt{3}) ,
\label{eq:defb}
\end{equation} 
see Figure \ref{hexagonal}.  The {two} sub-lattices are shifted with
respect to {each} another by {the vector} ${\bm a}_2 =
(\bm{b}_1+\bm{b}_2)/2$.  This representation is useful when evaluating
Coulomb lattice sums (see Section \ref{analytic-approach}).

Let $S$ denote the surface of a finite section of the 2D Wigner
crystal formed by $N$ particles.  We shall take the limit $S$ (and
thus $N$) $\to \infty$, and the electro-neutrality condition imposes
that the charge density, $\sigma e$, is given by
\begin{equation}
{\sigma = \frac{N}{S}} .
\end{equation}
There is {exactly} one particle per rhombus of surface $\sqrt{3}a^2/2$; {thus}
\begin{equation}
\frac{S}{N} = \frac{\sqrt{3}}{2} a^2, \qquad \mbox{i.e.} \qquad
\frac{\sqrt{3}}{2} a^2 \sigma = 1 .
\end{equation}

The Coulomb interaction energy of two particles {separated by a}
distance $r$ is {given by} $e^2/r$.  The ground-state energy, {$E_0$,}
of {an infinitely large system (consisting of the hexagonally arranged
  particles and the neutralising background) is found to be}
\cite{Bonsall77}
\begin{equation} 
\label{energy0}
\lim_{N\to\infty} \frac{E_0}{N} = \frac{1}{2} \sum_{i,j=-\infty\atop (i,j)\ne (0,0)}^{\infty} \frac{e^2}{\sqrt{(ai+\frac{1}{2}aj)^2+(\frac{\sqrt{3}}{2}aj)^2}} - \mbox{background} = c e^2 \sqrt{\sigma} ,
\end{equation}
the prefactor $c=-1.960515789319\ldots$ being known as the Madelung constant.

\subsection{The ``in''- and ``out''-branches}
\label{ssec:inandout}

\begin{figure}[htb]
\begin{center}
\includegraphics[width=.8\linewidth]{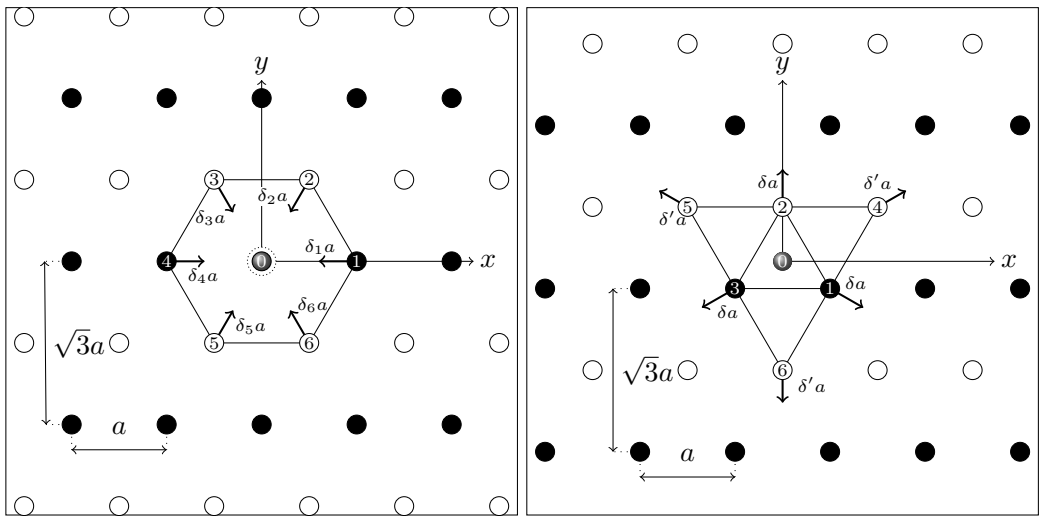} 
\end{center}
\caption{Typical configurations along the two branches, together with
  labeling of neighbours (from 1 to 6) and their displacements. The
  tagged charge carries the label 0 and its projection is shown in
  grey.  Left panel: schematic picture along the
  ``out''-branch. Arrows indicate possible displacements ($\delta_1 a$
  to $\delta_6 a$) of particles 1 to 6, induced by the removal of the
  tagged particle. Right panel: same, along the ``in''-branch. Arrows
  indicate possible displacements ($\delta a$ for the nearest
  neighbours and $\delta' a$ for the next-nearest neighbours) of
  particles 1 to 6, induced by the presence of the tagged particle.}
\label{schemecoord1}
\end{figure}

As alluded to above, locally stable configurations may be found for a
given $d$, which in turn complicates the search
  for the ground-state of the system.  These states seem to be
remnants of the coordination six structure valid at $d=0$ on the one
hand, and of the coordination three structure valid at large $d$ on
the other hand. To circumvent the ensuing metastability problem, we
have defined -- and investigated separately -- two branches for
computing the energies.  (i) Along the ``out''-branch, {we take our
  tagged particle (extruder)} from its {hexagonally coordinated}
position in the perfect Wigner crystal, {and place it at a distance
  $d$, letting then $d$ range from $0$ to $\infty$}.  {(ii) Along the}
``in''-branch, the {tagged} particle (intruder), located at
''$d\to\infty$'' {is placed ''above''} the center of an initially
undistorted triangle {formed by particles in the ideal 2D Wigner
  crystal, i.e., we gradually decrease the value of $d$ from $\infty$
  to 0.}  The corresponding energies of the {entire} system (``Wigner
crystal + tagged particle'') are denoted by $E^{\rm out}(d)$ and
$E^{\rm in}(d)$, respectively.  Figure \ref{schemecoord1} displays
typical configurations for the two branches, and defines variables
 that will be measured subsequently for quantitative
analysis.

To obtain meaningful results for large system sizes, we subtract from
$E^{\rm out}(d)$ and $E^{\rm in}(d)$ the ground-state energy of the
{\it perfect} Wigner crystal, $E_0$, [cf. Equation (\ref{energy0})]
when the tagged particle is still part of the Wigner crystal. We
thereby define
\begin{equation} 
\label{delta_e}
\delta E^{\rm out}(d) = E^{\rm out}(d)-E_0 \quad , \quad
\delta E^{\rm in}(d) = E^{\rm in}(d)-E_0 .  
\end{equation} 
Our main interest focuses on how $\delta E^{\rm out}(d)$ and $\delta
E^{\rm in}(d)$ vary as functions of $d$.  For a given value of $d$,
the state with the lower energy is considered as the ground-state,
while the other one is metastable. It will be shown that a transition
takes place between the two branches at a distance $d_t\neq 0$ which
is determined by the equality $\delta E^{\rm out}(d_t) = \delta E^{\rm
  in}(d_t)$.

Before embarking on a detailed study, it is useful to work out the
limiting values of $\delta E^{\rm out}(d)$ and $\delta E^{\rm in}(d)$, i.e. for $d \to 0$ and $d \to \infty$.  We have $\delta
E^{\rm out}(0) = 0$ by definition. On the other hand, $\delta E^{\rm
  in}(0)$ may differ from zero, if the system, along the
``in''-branch, remains trapped in a metastable state, even at
$d=0$. In that case, we expect $\delta E^{\rm in}(0) >0$ for
consistency.  Considering next the asymptotic value $\lim_{d \to
  \infty} \delta E^{\rm in}(d)$, we note that the interaction of the
tagged particle with the Wigner crystal vanishes as $d$ tends towards
infinity. In this situation, the remaining $(N-1)$ particles form a
perfect Wigner crystal with a lattice spacing $b$ given by
\begin{equation} 
\frac{S}{N-1} = \frac{\sqrt{3}}{2} b^2 \qquad
\mbox{i.e.} \qquad \frac{\sqrt{3}}{2} b^2 \sigma \frac{N-1}{N} = 1 .
\end{equation} 
The total energy of the $(N-1)$ charges forming a Wigner crystal of
spacing $b$ is proportional to $(N-1) e^2/b$, such that
\begin{equation}
E^{\rm in}(\infty) = c e^2 \sqrt{\sigma} N \left( \frac{N-1}{N}
\right)^{3/2} .  
\end{equation} 
Thus, 
\begin{equation} 
\label{asymp}
\delta E^{\rm in}(\infty) = c e^2 \sqrt{\sigma} N \left[ \left( \frac{N-1}{N} \right)^{3/2} - 1 \right] \mathop{\sim}_{N\to\infty} - \frac{3}{2} c e^2 \sqrt{\sigma} \, > \, 0 .  
\end{equation} 
At first sight, the prefactor $3/2$ {in the above relation} is
counter-intuitive: one would {rather} expect {a factor of one}, since
removing one particle from the Wigner crystal increases the {total}
energy by $-c e^2 \sqrt{\sigma}$ [see Equation (\ref{energy0})].
{However,} the point is that an infinitesimal increase of the lattice
spacing (after {having taken} off one particle from the system), when
multiplied by an (infinite) $N$, contributes to the energy change by a
finite amount. Note that a similar phenomenon would hold for any
inverse-power-law potential. Finally, the value of $\delta E^{\rm
  out}(\infty)$ might differ from {the expression given in Equation}
(\ref{asymp}), provided that the ``out''-branch is frozen in a local
minimum for $d \to \infty$.  We should nevertheless observe that
$\delta E^{\rm out}(\infty) > \delta E^{\rm in}(\infty) = -(3/2) c e^2
\sqrt{\sigma}$.

Finally, instead of the distance $d$, we will use in the following the
dimensionless distance $\eta$ defined by
\begin{equation} 
\eta = d \sqrt{\frac{\sigma}{2}} , \quad \hbox{so that} \quad \left( \frac{d}{a} \right)^2 \,=\, \sqrt{3} \,\eta^2 .  
\end{equation}

\section{Numerical approaches} 
\label{numerical-approach}

The problem, as specified in the Introduction, involves an infinite
monolayer of charged particles on a neutralising background, with a
test particle held fixed at a given vertical distance $d$ from the
monolayer. It is as such not amenable to numerical treatment.  For the
sake of numerical implementation, we shall consider a finite section
of the monolayer, and impose periodic boundary conditions in the $(x,
y)$-plane, a routine practice, thereby replicating the cell that
contains the section in question and the tagged charge. Keeping in
mind that we are dealing with long range interactions, finite size
effects must be carefully studied, in order to guarantee that the
observations made are not a consequence of the finiteness of the
setup.  Due to periodic replication, the system under scrutiny becomes
a bilayer, with inter-layer spacing $d$, number density
$\rho_1$ close to $\sigma$ on the ``bottom'' layer, and a {\it finite}
although small number density of particles on the ``top'' layer,
$\rho_2$:
\begin{equation} 
-\rho_1+\sigma   =  \sigma \left(-\frac{N-1}{N}+1\right) =  \sigma \frac{1}{N} \ , \qquad
-\rho_2  =  -\sigma \frac{1}{N} .  
\end{equation}
This means that for large separation $\eta$, the system behaves as a
capacitor with surface charges $\pm \sigma e/N$, and an energy $\delta
E(\eta)$ which consequently diverges like $\eta/N$, due to the the
finite electric field between the two plates.  This feature, which
sets in for $d \gg a\sqrt{N}$, however is immaterial here, since the
phenomena we shall study take place for $d$ on the order of the
lattice spacing $a$.

Two different numerical approaches were implemented: one is based on a
zero temperature energy minimisation technique, the other one on Monte
Carlo (MC) simulations at low temperature.

\subsection{Energy minimisation}
\label{energy-minimization}


To find the equilibrium configuration for a given dimensionless
distance $\eta$, we have to minimise the energies $\delta E^{\rm
  in}(\eta)$ and $\delta E^{\rm out}(\eta)$ -- cf. Equation
(\ref{delta_e}). Under the assumption of periodicity, we can calculate
these quantities by employing Ewald summation techniques
\cite{Mazars11}, which guarantee, with suitably chosen numerical
parameters, a relative accuracy of $10^{-5}$ or less. We chose cutoff
distances in real and reciprocal space to be
$R_{\rm{c}}=15/\sqrt{\sigma}$ and $K_{\rm{c}}=10\sqrt{\sigma}$,
respectively, and an Ewald summation parameter $\alpha=0.3$
\cite{Mazars11}.

We require an efficient gradient descent method to minimise the total
energy.  For this purpose we have employed the L-BFGS-B algorithm
\cite{Byrd95}. The derivatives of the energy with respect to the free
parameters (i.e., the positions of the particles) can be calculated
explicitly from the analytical expressions of the Ewald sums with high
numerical accuracy. A possible shortcoming of such a gradient descent
method is that the system can be trapped in local minima; this is the
reason why the ``out''- and ``in''-processes may lead to different
results. In practice, we have studied the ``out''-branch with a system
of $N=100$ charges, and its ``in''-counterpart with $N=101$ (in the
latter case, it is convenient to take $N$ as $n^2+1$, a square integer
plus one, since at large $d$, we are dealing with a perfect
undistorted crystal (with $n^2$ charges in the simulation cell), to
which the tagged particle should be added. The reason for choosing ensembles of this
  size is justified by the fact that for such a number of particles
  the interactions of a charge with its periodic images have become
  negligible.

We have verified our results by employing an optimisation tool based
on evolutionary algorithms
\cite{Holland75,Gottwald05,Doppelbauer10,Antlanger13} and have
compared the results. This more general approach does not require
following a particular branch.  Instead, the algorithm starts from
several random configurations. New configurations are created from one
or two existing ones and optimised using the L-BFGS-B method. This
process is repeated many times, combining good traits of previous
configurations and exploring new arrangements, until the excess energy
no longer improves.  In doing so, we recover as the optimal state one
of the configurations obtained following ``in''- or ``out''-branches,
depending on which one is more favourable. Therefore, this more
sophisticated method does not provide any energetic improvement over
the results that were obtained using the gradient-based approach with
suitable starting configurations, which furthermore yield an insight
on metastability.

\subsection{Monte Carlo simulations}
\label{simulations}

The Monte Carlo (MC) simulations reported here have been carried out
in the canonical ensemble with fixed $N$, a constant surface $S$, and
a finite temperature $T$.  The standard Metropolis algorithm has been
used throughout \cite{Newman:book:99}.  Periodic boundary conditions
were enforced (like in the energy minimisation route), and changes in
the shape of the simulation box were allowed. The long range nature of
the Coulomb interaction is again taken into account with the Ewald
summation technique, along similar lines as for previous studies on
Wigner bilayers \cite{Mazars:05b,Weis:01,Mazars:08}.

The one component plasma coupling constant for a two-dimensional
system is defined as $\Gamma=\sqrt{\pi\sigma}e^2/k_BT$.  Being
interested in ground-state properties, our goal is to know the $\Gamma
\to \infty$ behaviour of the system.  In the MC simulations, a
particularly high value was thus chosen, $\Gamma \simeq 1550$, about
ten times larger than the melting temperature
\cite{Monarkha:book:03,Clark:09}. This guarantees that the Wigner
monolayer, although investigated in MC at a non vanishing temperature,
is nevertheless in a crystalline state, with charges very close to
their ground-state positions. It should be kept in mind though that as
$d$ increases, the coupling energy between the reference charge and
the polarised crystal becomes weaker, and that the effect of a finite
temperature consequently becomes more prevalent. More specifically,
there exists an upper distance, diverging for small $T$ as $-\log T$,
  beyond which the
field created by the reference charge is insufficient to ``pin'' the
charges in the monolayer.  In order to gauge finite size effects (if
any), two system sizes have been considered: $N=2025$ and $N=256$.


One MC-cycle corresponds to a trial move of the $(N-1)$ mobile
particles and a trial change in the shape of the simulation box,
keeping the surface $S$ fixed. For ensembles with $N=2025$ particles,
$2\times 10^5$ MC-cycles have been performed in order to relax the
system from its initial condition; all ensemble averages, denoted in
the following by $ \langle .  \rangle$ have been taken during
$2-4\times 10^5$ additional MC-cycles.  For ensembles with $N=256$,
equilibration runs were carried out over $8\times 10^5$ MC-cycles and
averages were computed over $0.8-1.6\times 10^6$ MC-cycles.

\subsection{Localisation of particles, and structural properties}

In the energy minimisation approach, the relative displacements of the
particles with respect to their original positions can be easily
extracted, once the monolayer has relaxed and adapted to the presence
of the reference charge.  Of particular interest here are those
particles that are closest to the tagged charge (see schematic views
in both panels of Figure \ref{schemecoord1}).  There are therefore no
fluctuations in the particle positions, unlike in Monte Carlo, where
an accurate localisation of particles requires a somewhat more
elaborate analysis.

Particle positions are represented by 3D vectors,
$\bm{r}=\bm{s}+z\hat{\bm{e}}_z$ with $\bm{s}$ the in plane position
(perpendicular to $z$); for the tagged particle $\bm{s}=0$ and $z=d$
while for particles in the monolayer, $\bm{r}=\bm{s}$ and $z=0$.  To
describe on a quantitative level the structural properties of the
system, we have evaluated the pair correlation function between the
fixed reference particle and the other charges belonging to the
monolayer ($L$). This function depends on
$\eta$ and is defined as
\begin{equation} \label{Corr6}
\displaystyle g_{0}(s) = \frac{1}{2\pi s\sigma} \Big \langle \sum_{i\in L}\delta(s-\mid {\bm s}_{i}\mid) \Big \rangle ,
\end{equation} 
with $s=|{\bm s}|$. 

\begin{figure}[htb]
\begin{center}
\includegraphics[width=.8\linewidth,clip]{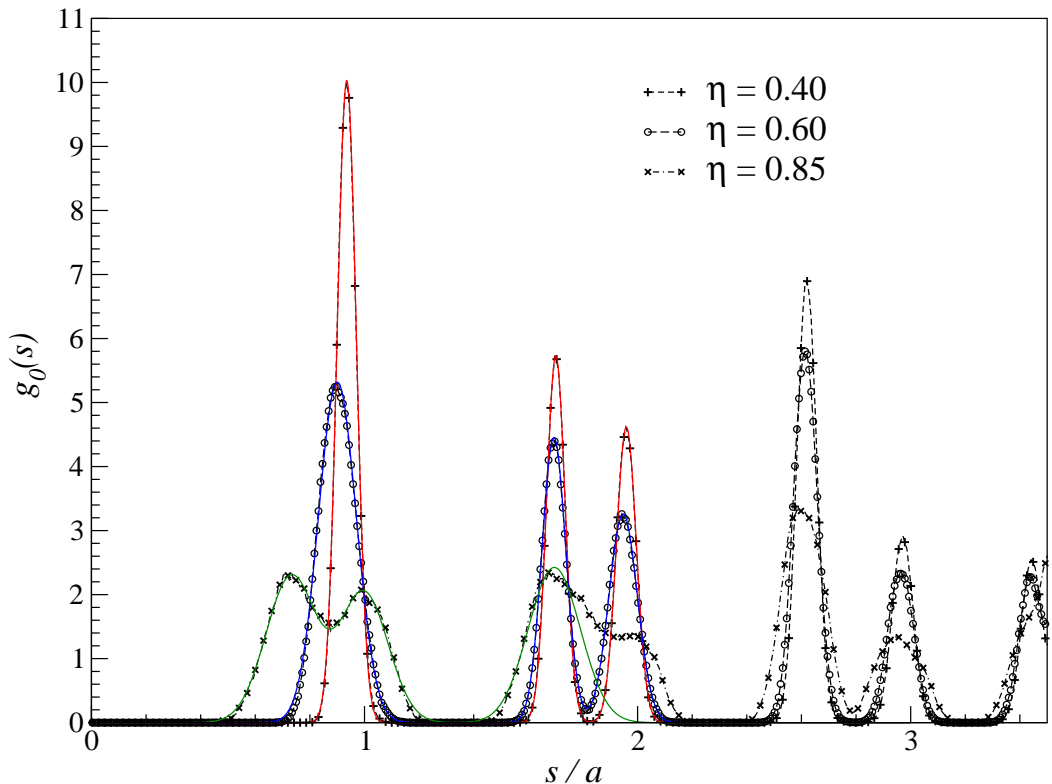}
\end{center}
\caption{(colors online) Pair correlation function $g_0(s)$ (defined
  in the text), as a function of the in-plane distance $s$ (in units
  of $a$) between particle 0 and the particles in the monolayer.
  Results are shown for three different values of $\eta$, along the
  ``out''-branch. Symbols: MC data; coloured lines: fits to the MC
  results using Equation (\ref{Corr-fit}) with $M_p=3$.}
\label{Fitg1}
\end{figure}


Since the coupling constant $\Gamma$ was chosen at a rather high value
-- $\Gamma \simeq 1550$ -- the use of the harmonic approximation for
the Wigner crystal is fully justified \cite{Ashcroft:book:76}. This
allows to approximate the $M_p$ first peaks in the correlation
functions as a sum of Gaussian functions:
\begin{equation}
\label{Corr-fit}  \displaystyle g_{0}(s) =\sum_{n=1}^{M_p} G_n^{(0)}\exp\left(-\frac{(s-s_n)^2}{2\lambda_n^2(\Gamma)}\right) \ , 
\end{equation} 
$s_n$ being the position of the $n$-th peak, $G_n^{(0)}$ its
amplitude, and $\lambda_n(\Gamma)$ its width.  Finally, the number of
particles $N_n$ that populate the $n$-th shell (which is defined by
the $n$-th peak) are computed via
\begin{equation}
\label{Nshell} 
\displaystyle N_n \, = \, 2\pi\sigma \, G_n^{(0)} \int_0^{\infty} s \, \exp\left(-\frac {(s-s_n)^2} {2\lambda_n^2(\Gamma)} \right) \,  ds .
\end{equation} 

\begin{figure}
\begin{center}
\includegraphics[width=.8\linewidth,clip]{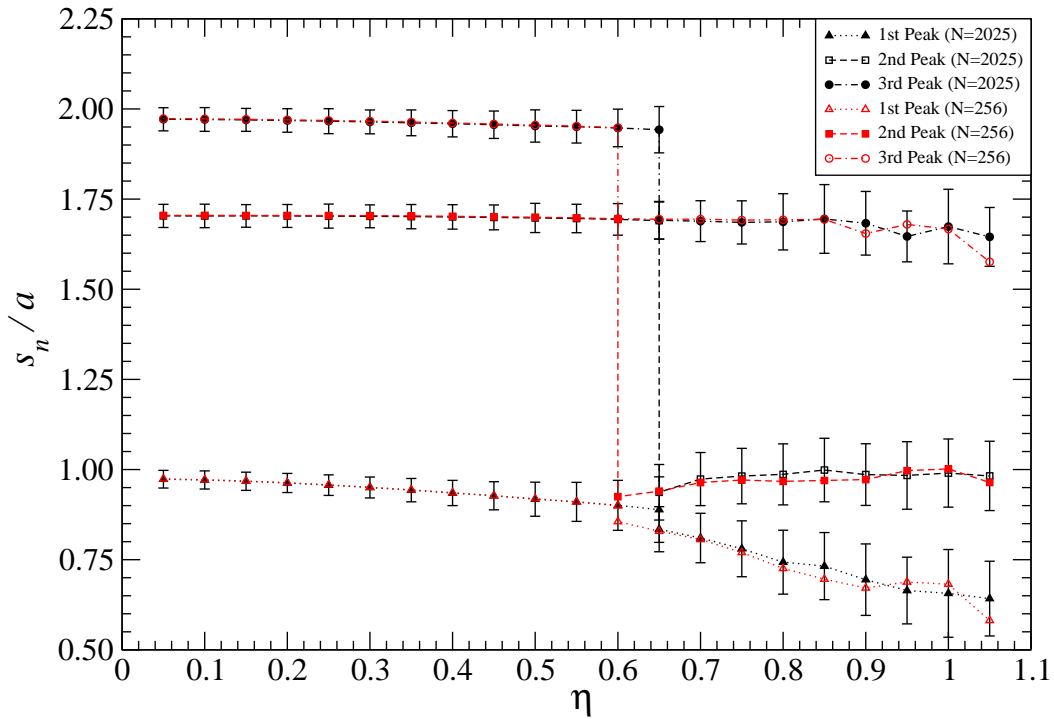}
\end{center}
\caption{Positions of the first three peaks ($s_1$, $s_2$, $s_3$) of
  $g_0(s)$ (shown in Figure \ref{Fitg1} for selected $\eta$-values) as
  functions of distance $\eta$, along the ``out''-branch. These MC
  results are shown for two different system sizes. Lines are guides
  to the eye. The error bars are the width of the Gaussians,
  $\lambda_n$, defined in Equation (\ref{Corr-fit}); they are shown
  only for $N= 2025$ particles since they are of comparable size for
  the 256 particle systems.
For $\eta=0$ the following values of the ideal 2D Wigner crystal are
quite well recovered: $s_1/a \simeq 1$, $s_2/a \simeq \sqrt{3}\simeq
1.73$ and $s_3/a \simeq 2$. }
\label{Fitpeaksg1}
\end{figure}

Examples for the correlation function $g_0(s)$ along the
``out''-branch are shown for three representative $\eta$-values in
Figure \ref{Fitg1}.
After fitting
by a sum of Gaussians -- cf. Equation (\ref{Corr-fit}) -- we can
determine the (average) positions of the particles within the
monolayer via the peak positions of the Gaussians; this information
allows, finally, for an accurate determination of the location (and
then the displacements) of the particles.
The positions of the first three peaks, $s_1$, $s_2$, and $s_3$, of
the correlation function $g_0(s)$, are shown in Figure
\ref{Fitpeaksg1} as functions of $\eta$.
As expected, for $\eta \to 0$ where the tagged particle is part of the
ideal 2D Wigner crystal, these three $s$-values tend to their ideal
undistorted hexagonal lattice expressions. These results show that
finite size effects are negligible, with very similar results for
small ($N=256$) as well as larger ($N=2025$) systems.

\section{Analytic treatment}
\label{analytic-approach}

In this Section, we aim at calculating analytically the energy
differences, $E^{\rm out}(\eta)$ and $E^{\rm in}(\eta)$ as defined in
Equation (\ref{delta_e}), along the ``out-'' and the
``in''-branches. For tractability, we assume that only the {nearest}
neighbours of the reference particle 0 can leave their {positions in
  the Wigner crystal}, while the remaining particles are assumed to be
fixed at their original positions in the lattice.  For the
``out''-branch, the rationale stems from Figure \ref{Fitpeaksg1},
showing that the second and third layers of neighbours are only weakly
displaced by the polarisation effect of the reference particle.  For
the ``out''-branch, this means that six particles are allowed to move
as $\eta$ changes, while for the ``in''-branch, there are only three charges (see below).  We use a recently proposed analytic
technique \cite{Samaj12a,Samaj12b}, which enables us to express
lattice Coulomb summations in terms of quickly convergent series of
the generalised Misra functions
\begin{equation} 
\label{Misra}
z_{\nu}(x,y) = \int_0^{1/\pi} \frac{{\rm d}t}{t^{\nu}}  {\rm e}^{-x t} {\rm e}^{-y/t} , \qquad y>0 .
\end{equation} 

\subsection{``Out''-branch}
\label{out-branch}

\begin{figure}
\begin{center}
\includegraphics[width=.3\linewidth,clip]{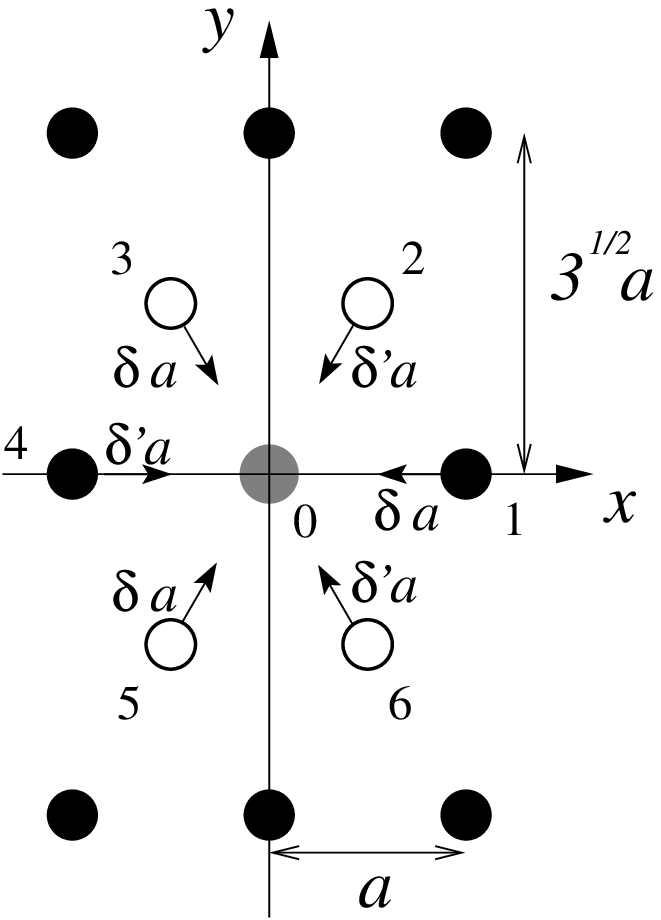}
~~~~~~~~~~~~~\includegraphics[width=.35\linewidth,clip]{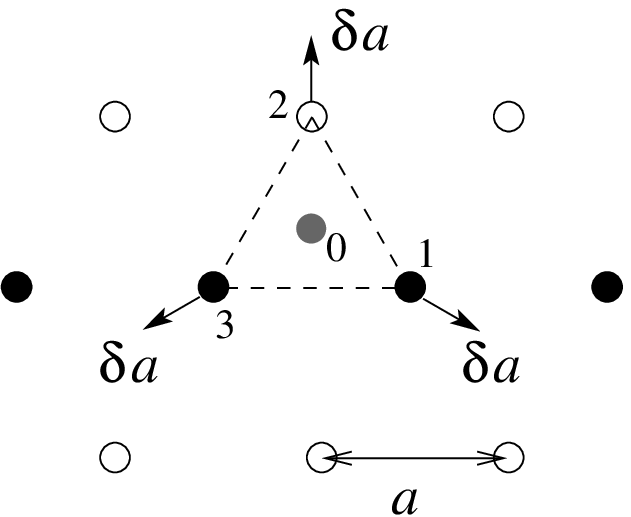}
\end{center}
\caption{Approximations used for the analytic treatment.  Left panel:
  along the ``out''-branch, only the six labeled neighbours are
  allowed to leave their perfect lattice positions.  Arrows indicate
  two possible, alternating displacements ($\delta a$ and $\delta
  'a$).  As compared to Figure \ref{schemecoord1}, we have
  $\delta_1=\delta_3=\delta_5=\delta$ and
  $\delta_2=\delta_4=\delta_6=\delta'$.  Left panel: along the
  ``in''-branch, only the three particles carrying indices 1 to 3 can
  move, along the arrows, when the tagged particle is approached.
In both ``in''- and ``out''-cases, the displacements
are radial.}
\label{schemecoord2}
\end{figure}

The left panel of Figure \ref{schemecoord2} defines those six
particles allowed to respond to the presence of the tagged particle
(the projection of which yields the shaded disk).  We have allowed --
by introducing parameters $\delta$ and $\delta'$ -- the possibility of
a symmetry breaking among neighbours.  The {hexagonal} Wigner {
  crystal} is represented by black and white {particles}, as discussed
above.  The {positions of the} fixed {black} particles are given by
{vectors} $a(j,\sqrt{3}k)$, where $j$ and $k$ are any two integers,
except for the pairs $(j,k)=(0,0), (1,0), (-1,0)$, the latter two
combinations corresponding to the black particles which are allowed to
be displaced.  The {positions of the} fixed {white} particles are
given by {vectors} $a\left( j+\frac{1}{2},
\sqrt{3}(k+\frac{1}{2})\right)$, where the pairs $(j,k)=(0,0), (-1,0),
(0,-1), (-1,-1)$ are excluded.

The energy of the given particle configuration, $E^{\rm
  out}(\eta;\delta,\delta')$, { has the following four contributions}:
\begin{itemize} 
\item[(i)] The interaction of particles labeled 1 to 6 with the
  {remaining} fixed {particles} and neutralising background:
\begin{eqnarray}
E_1(\delta,\delta') & = & 3 \frac{e^2}{a} \Bigg[  I(\delta,\sqrt{3}) + J(\delta,\sqrt{3}) + I(\delta',\sqrt{3})  + J(\delta',\sqrt{3}) \nonumber \\ & &
- \frac{1}{1-\delta} - \frac{1}{2-\delta} - \frac{2}{\sqrt{1-\delta+\delta^2}} - \frac{2}{\sqrt{3-3\delta+\delta^2}} \nonumber \\ & &
- \frac{1}{1-\delta'} - \frac{1}{2-\delta'}  - \frac{2}{\sqrt{1-\delta'+\delta'^2}} - \frac{2}{\sqrt{3-3\delta'+\delta'^2}} \Bigg] , \label{E1}
\end{eqnarray}
where the definitions of the lattice sums $I(\delta,\Delta)$,
$J(\delta,\Delta)$, and their series representations are given in
Equations (\ref{I}) and (\ref{J}) of the Appendix, respectively.  In
practice, {these series} must be {truncated} at some {finite value}
$M$, i.e. summed over the set $1,2,\ldots,M$ or $-M,-M+1,\ldots,M$.
Since the{se} series are quickly convergent, {we have chosen} here and
in what follows {a cutoff value} of $M=5$, which reproduces the exact
energy values to 17 decimal digits \cite{Samaj12b}.

\item[(ii)] The interaction of particles labeled 1 to 6 with each
  other:
\begin{equation}
E_2(\delta,\delta') = \frac{e^2}{a} \left[ \frac{6}{\sqrt{1-\delta-\delta'+\delta^2+{\delta'}^2-\delta\delta'}} +\frac{\sqrt{3}}{1-\delta} +\frac{\sqrt{3}}{1-\delta'} + \frac{3}{2-\delta-\delta'} \right] . 
\end{equation}

\item[(iii)] The interaction of the {tagged} particle 0 with the fixed
  {charges in the remaining Wigner crystal} and {with the
    neutralising} background
\begin{equation} 
\label{E3} 
E_3(\eta) = \frac{e^2}{a} \left[ K(\eta) + L(\eta) - \frac{6}{\sqrt{1+\sqrt{3}\eta^2}} \right] , 
\end{equation}
where the lattice sums $K(\eta)$ and $L(\eta)$ are defined in
Equations (\ref{K}) and (\ref{L}) of the Appendix, respectively.

\item[(iv)] The interaction of the tagged particle $0$ with particles
  labeled 1 to 6:
\begin{equation}
E_4(\eta;\delta,\delta') = 3 \frac{e^2}{a} \left[ \frac{1}{\sqrt{(1-\delta)^2+\sqrt{3}\eta^2}} + \frac{1}{\sqrt{(1-\delta')^2+\sqrt{3}\eta^2}} \right] .
\end{equation}
\end{itemize}

The total energy shift is {then} given by
\begin{equation}
\delta E^{\rm out}(\eta;\delta,\delta') = E^{\rm out}(\eta;\delta,\delta') - E^{\rm out}(0;0,0) ,
\end{equation} 
where $E^{\rm out}(\eta;\delta,\delta') = E_1(\delta,\delta') +
E_2(\delta,\delta') + E_3(\eta) + E_4(\eta;\delta,\delta')$.  For a
given {value of} $\eta$, the particle shifts $\delta$ and $\delta'$
are determined by minimising $\delta E^{\rm
  out}(\eta;\delta,\delta')$.  Within the present approximation,
{where we assume that only six nearest neighbours are allowed to
  move}, we obtain {throughout} $\delta=\delta'$ {irrespective of the}
value of $\eta$, i.e. symmetry breaking {does not take place}.

\subsection{``In''-branch}
\label{in-branch}

To study the ``in''-branch, we allow for the displacement of the three
neighbours represented in the right panel of Figure
\ref{schemecoord2}.
Assuming particle 3 to be located in the origin of the coordinate
system, the $x$- and $y$-coordinates of the tagged particle are given
by the vector ${\bm s} = a\left(1/2,\sqrt{3}/6\right)$. For a finite
  value of distance $\eta$, it is assumed that the positions of
  particles 1, 2 and 3 are shifted from their ideal positions by
  $\delta a$, in directions pointing away from particle 0.  The
remaining particles of the Wigner crystal are assumed to be fixed:
black particles at positions $a(j,\sqrt{3}k)$, excluding integer pairs
$(j,k)=(0,0), (1,0)$ and white particles at positions $a\left(
j+\frac{1}{2},\sqrt{3}(k+\frac{1}{2})\right)$, with $(j,k)\ne (0,0)$;
the aforementioned excluded pairs of indices correspond to the
positions of the mobile particles.

The energy {of the given particle configuration, $E^{\rm
    in}(\eta;\delta)$, bears the following four contributions:}
\begin{itemize}
\item[(i)] The interaction of particles labeled 1 to 3 with the
  {remaining, fixed particles in the Wigner crystal and with the}
  neutralising background:
\begin{equation}
\label{E1p}
E'_1(\delta) = 3 \frac{e^2}{a} \left\{ \frac{1}{\sqrt{3}} \left[ I\left(\frac{\delta}{\sqrt{3}},\frac{1}{\sqrt{3}}\right) + J\left(\frac{\delta}{\sqrt{3}},\frac{1}{\sqrt{3}}\right) \right] - \frac{2}{\sqrt{1+\sqrt{3}\delta+\delta^2}} \right\} . 
\end{equation}

\item[(ii)] The interaction of particles labeled 1 to 3 with each
  other:
\begin{equation} E'_2(\delta) = 3 \frac{e^2}{a} \frac{1}{1+\sqrt{3}\delta} .  
\end{equation}

\item[(iii)] The interaction of the tagged particle 0 with the fixed
  {charges in the remaining} Wigner crystal and {with the
    neutralising} background:
\begin{equation}
\label{E3p}
E'_3(\eta) = \frac{e^2}{a} \left\{ \frac{1}{2} \left[  \sqrt{3} K(\sqrt{3}\eta) - K(\eta) + \sqrt{3} L(\sqrt{3}\eta) - L(\eta) \right] - \frac{3}{\sqrt{\frac{1}{3}+\sqrt{3}\eta^2}} \right\} .
\end{equation}

\item[(iv)] The interaction of the tagged particle 0 with particles
  labeled 1 to 6:
\begin{equation}
E'_4(\eta,\delta) = 3 \frac{e^2}{a} \frac{1}{\sqrt{\left(\frac{1}{\sqrt{3}}+\delta\right)^2+\sqrt{3}\eta^2}} .
\end{equation}
\end{itemize}
Again, the lattice sums $I(\delta, \Delta)$, $J(\delta, \Delta)$,
$K(\delta, \Delta)$, and $L(\delta, \Delta)$ are given in the Appendix
in Equations (\ref{I}), (\ref{J}), (\ref{K}), and (\ref{L}).

The total shift {in energy is then} given by
\begin{equation}
\delta E^{\rm in}(\eta,\delta) = E'_1(\delta) + E'_2(\delta) + E'_3(\eta)  + E'_4(\eta,\delta) +  \frac{e^2}{a} \left( 3 - \frac{15}{\sqrt{2}3^{1/4}} c \right) ,
\end{equation} 
where the last (constant) term is determined by the asymptotic
condition (\ref{asymp}).  As before, the particle shift $\delta$ is
determined by minimising $\delta E^{\rm in}(\eta,\delta)$.

\section{Results}
\label{results}

\begin{figure}[htb]
\begin{center}
\includegraphics[width=0.8\textwidth,clip]{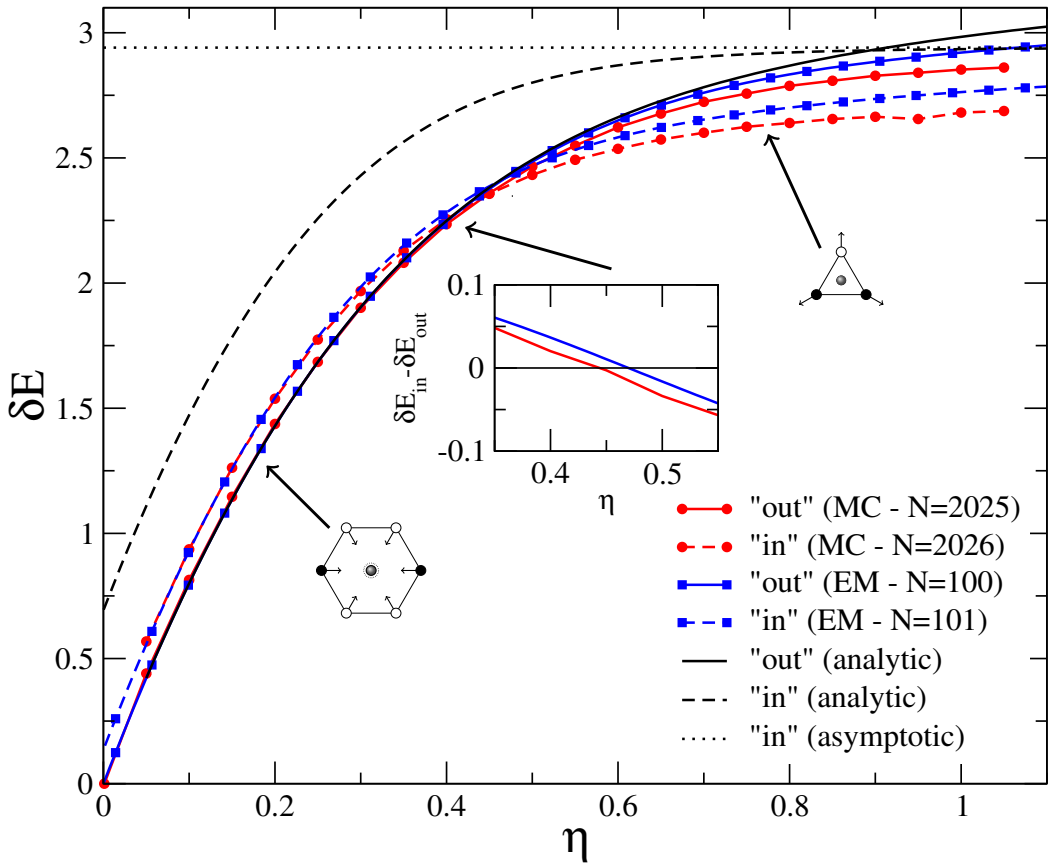}
\end{center}
\caption{(colors online) Excess system energy $\delta E$ with respect
  to the undistorted Wigner crystal, in units of $e^2 \sqrt{\sigma}$.
  Red, blue, and black lines are respectively for Monte Carlo (MC),
  energy minimisation (EM), and analytic results.  The full lines
  display the ``out''-branch data, while the broken ones are for the
  ``in''-branch.  Horizontal dotted line: prediction of Equation
  (\ref{asymp}) for the asymptotic energy along the ``in''-branch,
  $\delta E(\eta \to \infty)/(e^2 \sqrt{\sigma}) = -3c/2 \simeq
  2.94077$. The inset shows the difference between ``'in''- and
  ``out''-energies, for both MC and EM. }
\label{defectenergy} 
\end{figure}

The central object in our study is the total Coulombic energy of the
system (``monolayer + tagged charge''), suitably shifted by its value at $d=0$ to obtain
a well behaved quantity for large systems (with $N \gg 1$). Figure
\ref{defectenergy} therefore conveys our main results, showing $\delta
E^{\rm in}(\eta)$ and $\delta E^{\rm out}(\eta)$, as defined in
Equation (\ref{delta_e}), calculated via the two numerical approaches
as well as with the analytic method. First of all, the three methods
display very consistent results on the ``out''-branch. Only for large
$\eta$ do they start to depart, with the analytical prediction
providing expectedly higher energy configurations. This clearly stems
from the assumption that only the six labeled charges in
the left panel of Figure \ref{schemecoord2} are
mobile. Turning to the ``in''-branch, we observe that restricting the
mobile charges to now a set of three (see Section
\ref{analytic-approach}), becomes more problematic. The analytical
``in''-branch energy departs significantly from the results of energy
minimisation and MC, the latter two being again consistent (broken
lines).  We note that the limiting cases discussed in Section
\ref{ssec:inandout} are obeyed. The energy graph is indeed compatible
with the ``in''-bound (\ref{asymp}). In addition, we can extrapolate
from the data that $\delta E^{\rm out}(\infty) > \delta E^{\rm
  in}(\infty)$, as a fingerprint of metastability along the
``out''-branch, even at large $\eta$. On the other hand, at short
distances, a related metastable blocking is observed for the
``in''-branch, resulting in values of $\delta E^{\rm
  in}(0)$ that are slightly positive.


Within a good agreement between the two numerical methods, the ``in''-
and ``'out''-curves cross at $\eta_t \sim 0.47$, see also the inset of
Figure \ref{defectenergy}.  Thus, below this value, the stable state
lies on the ``out''-branch, while for $\eta \gtrsim 0.47$, the
``in''-branch structure is energetically more favourable. What are the
configurations of the charges along these two branches?
In particular, is coordination six retained along the
``out''-protocol, and likewise is coordination three retained all
along the ``in''-branch?  To answer these questions, we will focus
on the particle displacements.  Even though the ``out''- and the
``in''-branches are metastable for $\eta \gtrsim 0.47$ and for $\eta
\lesssim 0.47$, respectively, we will discuss some of their properties
not only for stable but also for metastable particle configurations,
as we have observed some interesting features.

\begin{figure}
\begin{center}
\includegraphics[width=0.9\textwidth,clip]{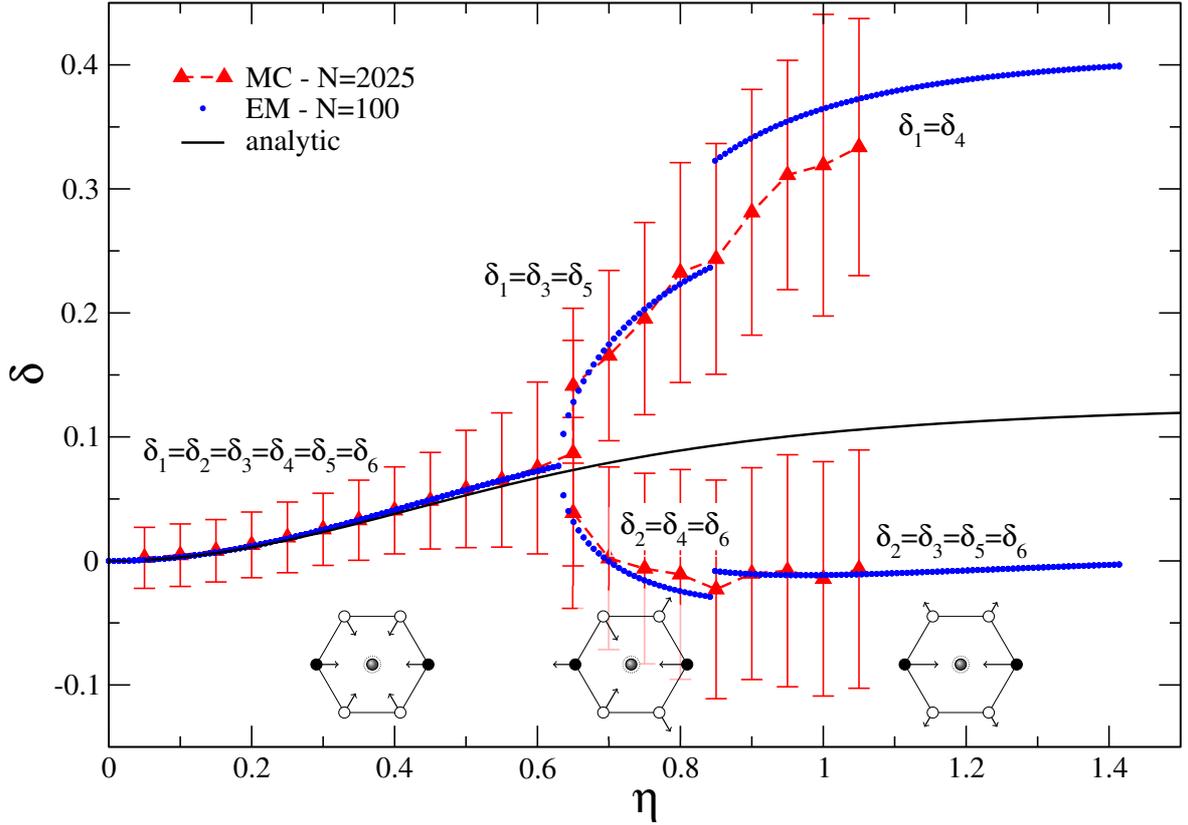}
\end{center}
\caption{(colors online) Deformations $\delta_1$ to $\delta_6$
  (in units of $a$) of the six nearest neighbours of the tagged
  particle along the ``out''-branch as defined in the left panel of
  Figure \ref{schemecoord1} as functions of $\eta$. Blue dots: results
  from the energy minimisation approach, black line: results from the
  analytic approach. Red symbols: deformations calculated from the
  positions of the first peak in $g_{0}(s)$; broken
  red line is drawn as a guide to the eye. Error bars indicate the
  $\sigma$-values of the Gaussian fit -- cf. Equation
  (\ref{Corr-fit}). The three schematic views provide a qualitative
  impression of the displacement of the six, most central particles.}
\label{outdeformation} 
\end{figure}

The displacement of the six most central particles ($\delta_1$ to
$\delta_6$), as calculated along the ``out''-branch, are accumulated
in Figure \ref{outdeformation} as functions of $\eta$. They are
computed from the particle positions at $\eta=0$, indicated by the
peaks of the corresponding $g_0(s)$.  For $0 \le \eta \lesssim 0.63$,
$\delta_i$-values ($i = 1, \cdots, 6$) are all positive and equal:
upon increasing $\eta$, the ring of inner particles contracts
uniformly towards the vacancy, while fully maintaining the six-fold
rotational symmetry of the particle arrangement (see left-most
schematic inset in Figure \ref{outdeformation}). However, when passing
this $\eta$-threshold value, symmetry breaking takes place: for $0.63 \lesssim \eta \lesssim 0.85$,
the particle configuration has now only three-fold rotational
symmetry. The six most central particles split up into two sets of
three: (i) particles of the first set (say 1, 3, 5), with their
$\delta_i$-values being throughout positive, have shifted towards
positions that are closer to the vacancy than the positions of the
second set; (ii) the displacements of the particles of the latter set
(say 2, 4, 6) decrease in this $\eta$-range monotonously; they even
become negative at $\eta \simeq 0.7$, indicating that for $\eta
\gtrsim 0.70$ particles of the second set are more distant from the
vacancy than in the ideal Wigner crystal. At this point it
  should be noted, that the best and the second-best configurations as
  determined via energy minimisation often differ by minute
  differences in their energy (i.e., by a few tenth or less of a
  percent).
  
  On the other hand, the analytical treatment does not predict this very scenario, even if it {\it a
  priori} allows for a symmetry breakdown of the type ``3 + 3'' (see
Figure \ref{schemecoord2}). This points to the subtle
effect of charge displacement beyond the first ring of neighbours,
which although small, can influence the structure in a non-trivial
way.  We emphasise at this point the excellent agreement between data
obtained from the energy minimisation approach and results extracted
from MC simulations, that extends for distances of the tagged particle
up to $\eta \simeq 0.85$. Finally, as we pass this $\eta$-value,
another transition takes place to a configuration with only two-fold
symmetry (see the right-most particle sketch of Figure
\ref{outdeformation}): two sets of particles form with two particles
(say, indices 1 and 4), the other one with four particles (say,
indices 2, 3, 5, and 6). According to the minimisation approach, this
transition is discontinuous, while there are indications that in MC
simulations, this transition is continuous; due to the large error
bars (see discussion below) an unambiguous conclusion about the nature
of this transition is difficult to reach. For the former set of
particles (where the two charges are located on opposite positions
with respect to the vacancy), the $\delta$-values are positive. The
charges thus have approached the hole left by the tagged particle. The
other four charges (forming the second set) are characterised by
small, negative $\delta$-values (which, with increasing $\eta$, tend
to zero), indicating that these particles have moved away from the
central position of the vacancy.

At this point it is worthwhile to note that -- while the average
values for the displacement as computed via MC simulations are in 
close agreement with the data obtained with the energy minimisation
approach (and with the analytical results for $\eta \lesssim 0.65$) --
the fluctuations of the MC-data are non-negligible. To be more specific,
for $\eta \lesssim 0.6$, the agreement between the two sets of
numerical data is better than 1\% and at most 3 to 5 \% for higher
$\eta$-values (see data shown in Figure \ref{outdeformation}). However,
the error bars, given by the respective $\lambda_n$-values and
displayed in this figure are quite large, and do increase when $\eta$
increases. This is due to the fact that particles in the layer
are pinned by the external potential of the reference charge,
which becomes softer as $\eta$ grows. The resulting fluctuations
ensure that the MC algorithm explores thoroughly the phase space ``around''
the equilibrium positions; it is the excellent accuracy of the harmonic
approximation that is able to identify well the equilibrium
positions of the particles as compared with the positions obtained via
the energy minimisation approach and via the analytic method.
A similar analysis can be applied to the
energy curves shown in Figure \ref{defectenergy}.

\begin{figure}
\begin{center}
\includegraphics[width=.8\linewidth,clip]{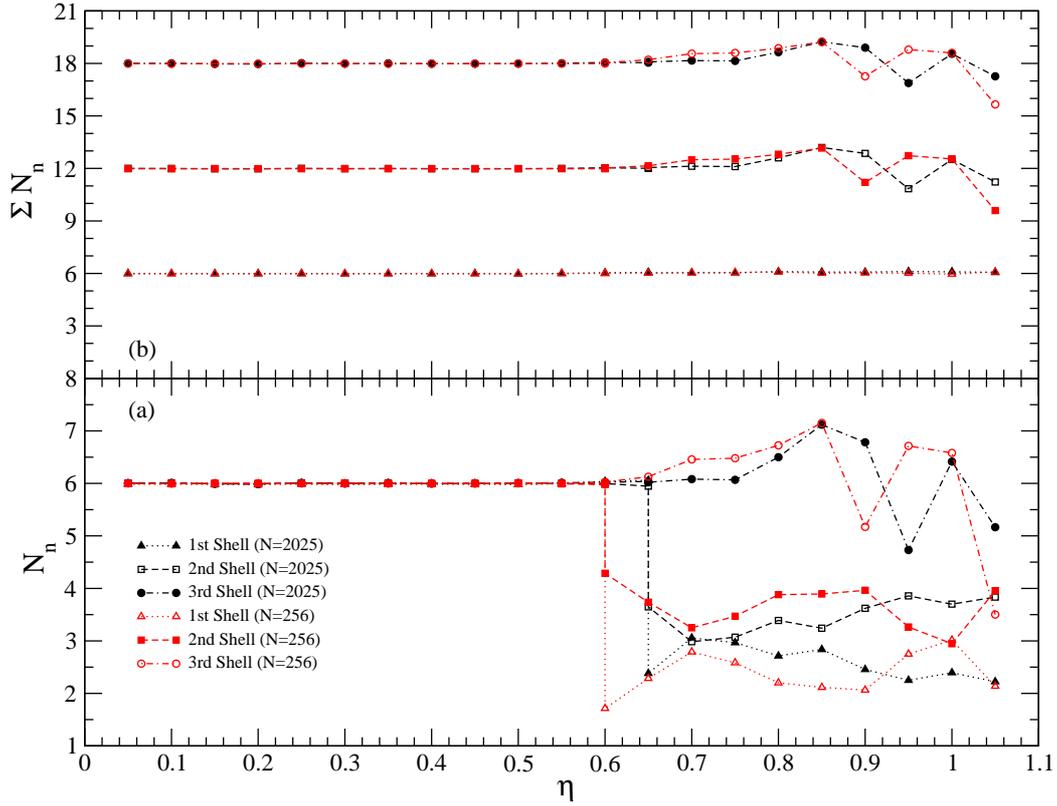}
\end{center}
\caption{MC simulations along the ``out''-branch.  Bottom panel:
  number of neighbours, $N_n$, in the first three shells ($n = 1, 2,
  3$) as functions of $\eta$.  Top panel: sum over the neighbours
  located {\it within} the first, the second, and the third shell, as
  functions of $\eta$. Results are shown for two different system
  sizes, lines are drawn as guide to the eye. }
\label{Fitneighg1}
\end{figure}

We now proceed with a more quantitative analysis of the aforementioned
splitting of the six most central particles in two subsets of
charges. This fact is visible in the $s_1$-branch shown
in Figure \ref{Fitpeaksg1}, and also appears in the bottom panel of
Figure \ref{Fitneighg1}.  The other two branches of Figure
\ref{Fitpeaksg1} (displaying data for $s_2$ and $s_3$) show a weak,
monotonous decrease with increasing $\eta$, indicating that not only
the nearest neighbours are affected by the displacement of the tagged
particle, but also second and third nearest neighbours (even though
their displacements are much weaker). 
Finally, the top panel of Figure \ref{Fitneighg1} provides evidence
that the number of particles in each shell assumes an essentially
constant and $\eta$-independent value of six (deviations for $\eta
\gtrsim 0.7$ can be attributed to the statistical noise in MC
simulations). We conclude that the
  perturbation induced by the tagged charge progresses shell-by-shell
  throughout the crystal.

\begin{figure}
\begin{center}
\includegraphics[width=.8\linewidth,clip]{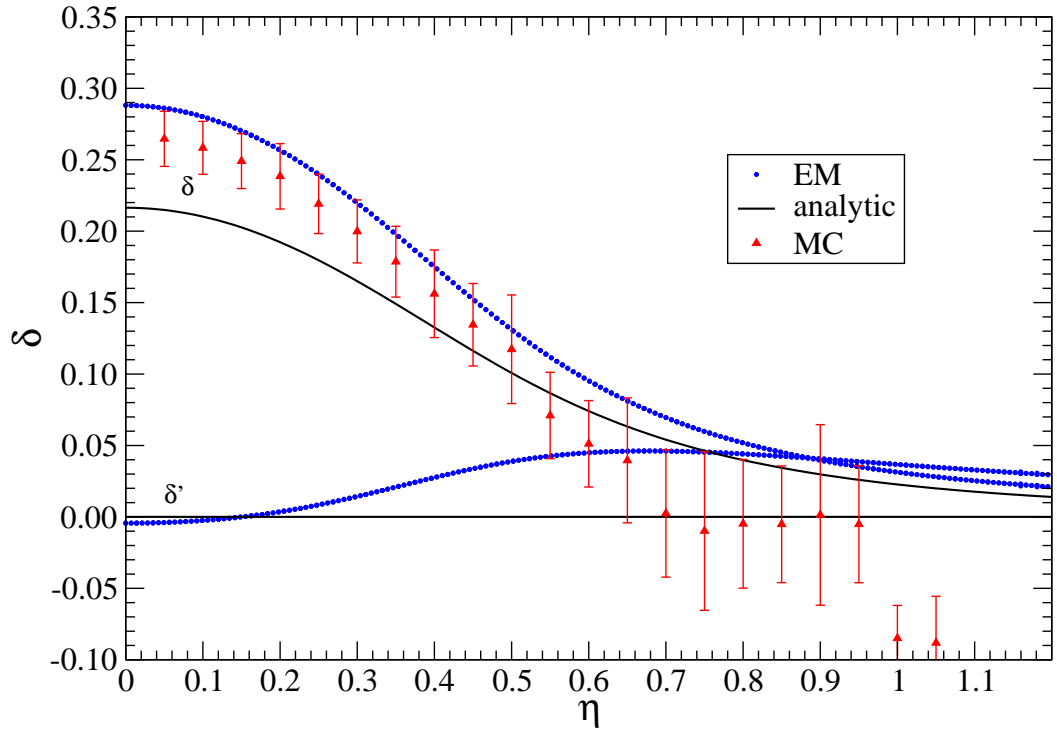}
\end{center}
\caption{Same as Figure \ref{outdeformation}, along the
  ``in''-branch. The displacements (defined in the right
    panel of Figure \ref{schemecoord1}) are computed from the
  positions occupied at large $\eta$. The energy minimisation (EM) was
  performed on a system with $N=101$ particles, while $N=2026$ in MC. }
\label{Energy_out}
\end{figure}

To complete our discussion on the displacement of the particles, we
turn to the ``in''-branch: here the scenario is simpler since
three-fold symmetry is maintained along the entire branch, with the
$\delta$-values decreasing continuously, to vanish at large $\eta$.
Since the analytical prediction is less reliable here, as explained
earlier, less effort was devoted to studying that branch.  Only
nearest neighbour results from MC simulations are shown,
since the localisation of charges is less accurate for the next
nearest neighbours (i.e., $\delta'$ data). It can be seen
in Figure \ref{Energy_out} that the trend found in MC, EM and
analytically is consistent, and that the small differences between the
two numerical data sets do not alter the good agreement found at the level of the
energy, see Figure \ref{defectenergy}. 
The energy minimimization route allows for an
  accurate determination of the individual displacements $\delta$ and
  $\delta'$ of each of the labeled particles in the right panel of
  Figure \ref{schemecoord1}; we find that particles carrying labels 1,
  2, and 3 are displaced (within numerical accuracy) by the same
  displacement $\delta a$, while the other three particles (with indices
  4, 5, and 6) are shifted by the same $\delta 'a$-value. Thus we can
  conclude that throughout the entire ``in''-branch three-fold
  coordination is preserved.

\begin{figure}
\begin{center}
\includegraphics[width=.8\linewidth,clip]{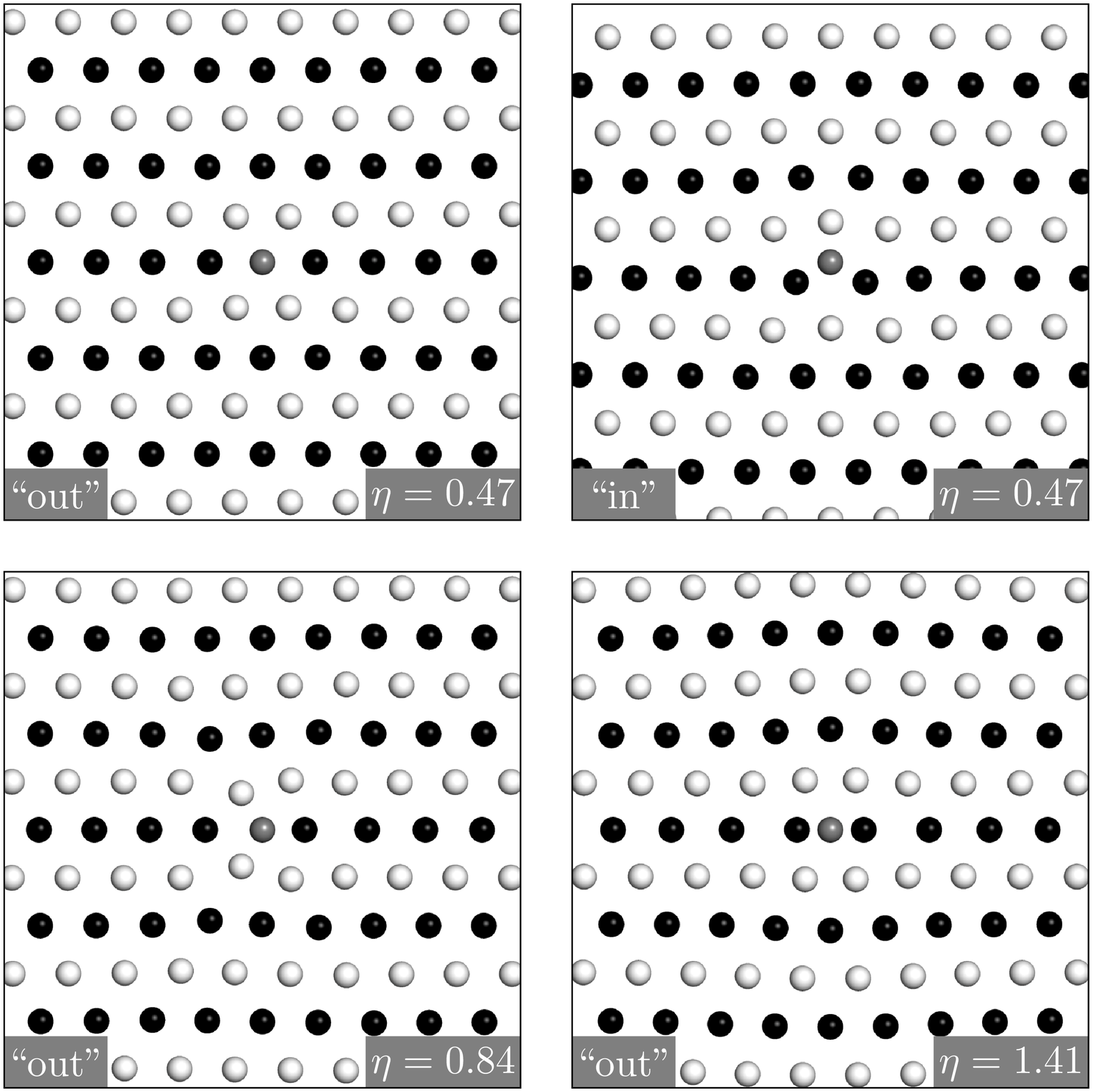}
\end{center}
\caption{Particle arrangements induced by the tagged charge (the projection of which is shown in grey, as 
in previous graphs). Top panels: the two competing configurations at the transition 
point $\eta_t \simeq 0.47$ (``in'' and ``out''  branches).
Bottom panels: two ``out''-branch configurations in the metastable region. }
\label{snapshots}
\end{figure}

Finally, in an effort to provide a {\it quantitative} impression
  about the influence of the tagged particle on the displacement of
  the charges in the monolayer, we have collected in Figure
  \ref{snapshots}  views of particle arrangements for
  selected $\eta$-values. In addition, we have concatenated sequences
  of equilibrium configurations for increasing (``out''-branch) and
  decreasing (``in''-branch) $\eta$-values into short animations, which
  are presented in the Supplementary Material. It can be seen in Fig.
  \ref{snapshots} that at the transition point $\eta_t$, the 
 like-energy configurations associated to the two branches are quite 
 distinct, and in addition do not exhibit significant displacements
 from their respective reference state ($d=0$ perfect hexagonal
 structure in the ``out'' case, and $d\to\infty$ perfect structure 
 in the ``in'' case, see Section \ref{ssec:inandout}). 
 The polarisation effect of the reference particle is thus quite 
 weak here, an information already conveyed in Figs. \ref{outdeformation}
 and \ref{Energy_out}. The bottom panel of Fig. 
 \ref{snapshots} illustrates the symmetry breakdown phenomenon
 which arises, along the ``out'' branch, for $\eta>0.63$.
The coordination number, which has a value of six for $\eta<0.63$,
decays to three (as is seen for $\eta=0.84$), and ultimately to two
for larger distances ($\eta=1.41$).

\section{Conclusion}
\label{conclusion}

We have investigated how a 2D Wigner crystal formed by
charges can be polarised, at $T=0$, by a single tagged charge.
Starting from a perfect crystal
on a planar neutralising background, we take off one reference charge
from its equilibrium position, and fix it at a given distance
$d$. This creates a vacancy in the plane, towards which
some of the other charges move. For all $d$ values, we have determined
the ground-state properties and structure of the system.
Starting from the expected ground-states at $d=0$ and
$d\to\infty$ respectively, we followed the particle rearrangements
along the corresponding ``out''- and ``in''-branches, evidencing the
metastability of the system.  Two numerical techniques were used,
namely direct
energy minimisation and Monte Carlo simulations at a
sufficiently small temperature. A third, independent procedure, of the
evolutionary type, was also employed to check some of the results.  In
addition, an analytic treatment was performed, under the assumption
that only those neighbours that are closest to the tagged particle (in
either the $d=0$ or the $d\to\infty$ case) are allowed to move: thus
six along the ``out''-branch, and three along the ``in''-one.  All
point charges interact through the usual $1/r$ potential, which
requires careful numerical treatment, and use of recently derived
results for handling lattice summations with extremely high precision.
Since all particles bear the same charge and our interest is focused
on ground-state features, the results obtained are independent of the
value of the charge.

We proved the existence of a transition distance $d_t \simeq 0.47
\sqrt{2/\sigma}$, $\sigma$ being the background density, so that for
$d<d_t$, the reference charge is six-fold coordinated, while for
$d>d_t$, the coordination is three. In terms of lattice spacing $a$, the transition distance
reads $d_t \simeq 3^{1/4} \,0.47 a \simeq 0.62 \,a$.  The transition
between both states is discontinuous, and involves an energy barrier that has not been investigated
  in this contribution. The actual evaluation of the height of this
  barrier would require different tools than the ones used in this
  paper; we have therefore postponed this
  undoubtedly interesting question to a future contribution. One can
  assume (or anticipate) that this barrier is sufficiently high so
that once the system lies on a given branch (``in'' or ``out''), it
stays trapped in the corresponding energy valley: the ``out''-branch
was observed to be metastable for $d>d_t$, and conversely, the
``in''-branch is metastable for $d<d_t$.  While the ``out''-branch
state is of coordination six in its range of stability, a symmetry breaking takes place around $d
\simeq 0.63 \sqrt{2/\sigma}$, beyond which the ``out''-branch is
three-fold, and ultimately, two-fold coordinated.

As far as the ``out''-branch is concerned, good agreement between the
numerical and analytical results was reported. The situation
complicates for the ``in''-branch. First of all, the exact treatment
assumed {\it a priori} in that case only three mobile neighbours, a number which turned out to be insufficient. Allowing for more mobile
neighbours (say six, as for the ``out''-analysis), would certainly
improve the predictions. In addition, the ``in''-case is also more
elusive within the MC scheme. The reference state is indeed the large
$d$ case, where the coupling energy between the fixed
  reference charge and the monolayer becomes small, and is washed out
by the fluctuations induced by temperature. A more
refined study of the ``in''-branch is left for the future.  

An interesting
question, coupled to that of the energy barrier alluded to above, 
pertains to the structure
of the transition state (the one at the saddle point between the two states at
    $d_t$). It is left for future
investigations.

\section*{Acknowledgements}

It is a pleasure to dedicate this work to Pierre Turq,
who has made seminal contributions to Coulombic systems,
in particular ionic liquids and electrolyte solutions.  The authors
acknowledge financial support from the projects PHC-Amadeus-2012/13
(project number 26996UC) and Projekt Amad\'ee (project number FR
10/2012), from the Austrian Research Foundation FWF (project number
P23910-N16 and the SFB ViCoM, FWF-Spezialforschungsbereich F41) as
well as Grant VEGA No. 2/0049/12. The authors acknowledge also the
computation facilities (iDataPlex - IBM) provided by {\it Direction
  Informatique\/} of {\it Universit\'e Paris-Sud}. This work was
granted access to the HPC resources of IDRIS under the allocation
2013097008 made by GENCI.

\appendix
\section{Series representations of lattice sums}
\label{appendix}

The lattice sums for the energies (\ref{E1}), (\ref{E1p}), and their series representations in terms of the generalized Misra functions (\ref{Misra}), read as
\begin{eqnarray} 
I(\delta,\Delta) & = & \sum_{(j,k)\ne (0,0)}  \frac{1}{\sqrt{(j+\delta)^2+\Delta^2 k^2}} - \mbox{backgr}  = - \frac{4}{\sqrt{\Delta}} + \frac{1}{\sqrt{\pi\Delta}} \nonumber \\
& & \times \Bigg\{ 2 \sum_{j=1}^{\infty} \cos(2\pi j\delta) z_{3/2}(0,j^2\Delta) + 2 \sum_{j=1}^{\infty} z_{3/2}(0,j^2/\Delta) \nonumber \\
& & + 4 \sum_{j,k=1}^{\infty} \cos(2\pi j\delta) z_{3/2}(0,j^2\Delta+k^2/\Delta)  + \sum_{j=-\infty}^{\infty} z_{3/2}(0,(j+\delta)^2/\Delta) \nonumber \\ 
& & - \frac{\sqrt{\pi\Delta}}{\delta} + 2\sqrt{\pi} + 2 \sum_{j=1}^{\infty}  \sum_{k=-\infty}^{\infty} z_{3/2}(0,j^2\Delta+(k+\delta)^2/\Delta) \Bigg\}  \label{I} \\ J(\delta,\Delta) & = & \sum_{(j,k)\ne (0,0)}  \frac{1}{\sqrt{(j+\frac{1}{2}+\delta)^2+\Delta^2 (k+\frac{1}{2})^2}}  - \mbox{backgr} = - \frac{2}{\sqrt{\Delta}} + \frac{1}{\sqrt{\pi\Delta}}  \nonumber \\ 
& & \times \Bigg\{ 2 \sum_{j=1}^{\infty} (-1)^j \cos(2\pi j\delta) z_{3/2}(0,j^2\Delta) + 2 \sum_{j=1}^{\infty} (-1)^j z_{3/2}(0,j^2/\Delta) \nonumber \\ 
& & + 4 \sum_{j,k=1}^{\infty} (-1)^{j+k} \cos(2\pi j\delta)  z_{3/2}(0,j^2\Delta+k^2/\Delta) \nonumber \\ 
& &  + 2 \sum_{j=1}^{\infty} \sum_{k=-\infty}^{\infty}  z_{3/2}(0,(j-\textstyle{\frac{1}{2}})^2\Delta +(k-\textstyle{\frac{1}{2}}+\delta)^2/\Delta) \Bigg\} . \label{J}
\end{eqnarray}
The expression in (\ref{I}) is finite also in the limit $\delta\to 0$, due to the fact that
\begin{equation}
z_{3/2}(0,\delta^2/\Delta) \mathop{\sim}_{\delta\to 0}  \frac{\sqrt{\pi\Delta}}{\delta} - 2\sqrt{\pi} + O(\delta) .
\end{equation}

The lattice sums for the energies (\ref{E3}) and (\ref{E3p}) are given by
\begin{eqnarray}
K(\eta) & = & \sum_{(j,k)\ne (0,0)} \frac{1}{\sqrt{j^2+3 k^2+\sqrt{3}\eta^2}} - \mbox{backgr} \nonumber \\ 
& = & \frac{1}{\sqrt{\pi} 3^{1/4}} \left[ I_3((\pi\eta)^2,0) + I_3(0,\eta^2) \right] , \label{K} \\
L(\eta) & = & \sum_{(j,k)}  \frac{1}{\sqrt{(j+\frac{1}{2})^2+3 (k+\frac{1}{2})^2+\sqrt{3}\eta^2}} - \mbox{backgr} \nonumber \\ 
& = & \frac{1}{\sqrt{\pi} 3^{1/4}} \left[ I_2((\pi\eta)^2,0) + I_4(0,\eta^2) \right]. 
\label{L}
\end{eqnarray}
The definitions of the integrals over the Jacobi theta functions $I_2$, $I_3$ and $I_4$ are presented respectively in Equations (61), (62) and (63) of Ref. \cite{Samaj12b}.

\end{document}